\documentclass[3p,times,procedia]{elsarticle}
\usepackage{ecrc}
\volume{00}

\firstpage{1}

\journalname{Transportation Research Procedia}

\runauth{Vianey Darsel}

\jid{trpro}

\usepackage{amssymb}
\usepackage{amsmath}
\usepackage{algorithm}
\usepackage{algpseudocode}
\usepackage{bbm}
\usepackage{svg}
\usepackage{multicol}

\biboptions{authoryear}

\usepackage[figuresright]{rotating}
\usepackage[bookmarks=false]{hyperref}
    \hypersetup{colorlinks,
      linkcolor=blue,
      citecolor=blue,
      urlcolor=blue}

\begin{document}
\begin{frontmatter}



\dochead{World Conference on Transport Research - WCTR 2026 Toulouse 6-10 July 2026}%

\title{THE FRAGILITY OF SYNTHETIC POPULATIONS OVER TIME: AN EVALUATION OF FORECASTING STRATEGIES FOR AGENT-BASED MODELS}

% The Fragility of Synthetic Populations Over Time: An Evaluation of Forecasting Strategies for Agent-Based Models
% Previous: AGENT-FORECASTING METHODS FOR SIMULATION IN THE FUTURE


\author[a]{Vianey Darsel\corref{cor1}} 
\author[a]{Etienne Côme}
\author[b]{Jeppe Rich}
\author[b]{Francisco Camara Pereira}
\author[a]{Latifa Oukhellou}

\address[a]{COSYS-GRETTIA, Université Gustave Eiffel, Champs-sur-Marne, France}
\address[b]{Danmarks Tekniske Universitet, Kongens Lyngby, Denmark}

\begin{abstract}

Agent-based simulation models are widely used to assess the future mobility projects. However, their effectiveness hinges on the accuracy of synthetic populations, which are often generated using data from a single reference year. This study explores the temporal validity of these populations by evaluating their performance over multi-year datasets in a simulation-like environment. We focus on multi-dimensional agents, ensuring that all attributes critical to the simulation are comprehensively represented. To assess the quality of synthetic populations, we compare three generation methods: direct use of original training data, Bayesian Network sampling, and diffusion model. We then evaluate four agent-forecasting approaches: static projection (both idealistic and realistic), resampling, and dynamic projection. While dynamic projection holds potential, it requires complex longitudinal data to model the evolution of agent attributes. To address this, we introduce a heuristic model based on aggregated statistics, though its limitation lies in ignoring attribute correlations, which reduces its accuracy.
Our results demonstrate that synthetic populations lose relevance over time, emphasizing the necessity of effective agent-forecasting. Among the methods tested, idealistic static projection yields the best performance but requires data that is difficult to obtain. The other methods show limited improvements, highlighting the ongoing challenges in accurately forecasting agent attributes.

\end{abstract}

\begin{keyword}
Agent-forecasting; Population Synthesis; Agent-based Simulation; Deep Generative Models




\end{keyword}
\cortext[cor1]{Corresponding author. Tel.: +33.1.81.66.86.79.}
\end{frontmatter}

%\correspondingauthor[*]{Corresponding author. Tel.: +0-000-000-0000 ; fax: +0-000-000-0000.}
\email{vianey.darsel@univ-eiffel.fr}

% \linenumbers

% \vspace*{-12pt}
\clearpage
\begin{nomenclature}
\begin{deflist}[A]
\defitem{${}^y\mathbf{X}$}\defterm{Multi-dimensional distribution of the population for year $y$}
\defitem{${}^y{X}_k$}\defterm{Distribution of the $k$-th attribute of ${}^y\mathbf{X}$}
\defitem{${}^y\mathbf{S}^n$}\defterm{$n$ realisations of a random variable following the distribution of ${}^y\mathbf{X}$}
\defitem{${}^y\mathbf{s}^i$}\defterm{The $i$-th realisation of a random variable following the distribution of ${}^y\mathbf{X}$}
\defitem{${}^y\tilde{\mathbf{S}}$}\defterm{A set of samples approximating a set of realisations from ${}^y\mathbf{X}$}
\defitem{$N_y$}\defterm{Target size for ${}^y\tilde{\mathbf{S}}$}
\defitem{$\Omega_k$}\defterm{Sample space for ${}^y{X}_k$. It corresponds to all modalities that ${}^y{X}_k$ could take over all years.}
\defitem{$\omega_k$}\defterm{A value from $\Omega_k$.}
\defitem{$f_k(\mathbf{S},\omega_k)$}\defterm{\hspace*{0.6cm}The observed frequency of the modality $\omega_k$ in the set $\mathbf{S}$.}
\defitem{${}^yu_j$}\defterm{An updating rule in the dynamic projection model.}
\defitem{$g(\mathbf{S})$}\defterm{Function associated to a generative model to generate a synthetic population with similar characteristic than the input population $\mathbf{S}$.}

\end{deflist}
\end{nomenclature}\vspace*{6pt}

%% START TEXT %%%%
%\enlargethispage{-7mm}

\section{Introduction}
\label{introduction}

Agent-based simulation models are powerful tools for estimating population flows within a geographic area. These models rely on a synthetic population that represents the total population, generated from limited datasets. While population synthesis has been extensively studied for reconstructing populations at a fixed point in time (typically the year of the original data), this article addresses a less explored challenge: agent-forecasting, or generating populations for future scenarios.

Agent-based simulations are particularly effective for analyzing mobility scenarios, whether based on current conditions or hypothetical settings. They play a critical role in the early stages of transportation projects, such as new transit lines, infrastructure development, or policy interventions, by enabling impact assessment and prediction. However, such projects often span decades: for example, Copenhagen’s Inderhavnsbroen bridge took five years to complete\footnote{\href{https://da.wikipedia.org/wiki/Inderhavnsbroen}{Inderhavnsbroen - Wikipedia}}, while the Grand Paris Express metro lines are expected to require 13 to 20 years\footnote{\href{https://fr.wikipedia.org/wiki/Grand_Paris_Express}{Grand Paris Express - Wikipedia}}. This long-term nature underscores the need for robust agent-forecasting methods.

Agent-forecasting shares conceptual links with demographic forecasting, which projects future population trends at an aggregate level \citep{booth_demographic_2006}. While demographic forecasts provide macro-level insights (e.g., population size, age distribution, and geographic spread), agent-forecasting operates at a microscopic level, generating individual-level attributes. Although national-level forecasting methods are well-documented, small-area projections remain underdeveloped \citep{wilson_methods_2022}. Agent-forecasting can bridge this gap by deriving aggregated statistics from synthetic populations, offering granularity for both public and private sectors to anticipate consumption patterns and policy needs \citep{mazzuco_developments_2020}.

% Agent-forecasting has some links with demography, especially with population forecasts, which explore the future demographic trends \citep{booth_demographic_2006}. Population forecasts aim to collect aggregate information, whereas agent-forecasting is at a microscopic level. In population forecasts, while there is extensive literature on national-level forecasting methods, \citet{wilson_methods_2022} note a gap for small-area projections. The authors explain that agent-forecasting can be used to that extend, as aggregated statistics can then be derived based on the attributes of the synthetic population.
% Population forecasts encompass various aspects such as population size, age distribution, and geographical spread. Such aggregated indicators are valuable for stakeholders, enabling them to plan for demographic shifts and tailor policies to meet the needs of future inhabitants. Additionally, both public and private sectors can utilize these forecasts to anticipate consumption patterns and address the evolving needs of the population \citep{mazzuco_developments_2020}. Conversely, the aggregated statistics from population forecasts can be used to help fitting agent-forecasting models.
 
Unlike population synthesis, agent-forecasting has received limited attention in the literature. There are three primary methodological approaches. 
The first two approaches (static projection and resampling) replicate samples from the initial dataset to satisfy statistical constraints for the target year, either on the marginals or on the distribution of selected attributes. These methods depend on the quality of the initial set, which becomes less relevant as time goes by. In this work, we investigate the impact of the interval between the training and testing data on the quality of the population, its representativeness, and its realism. 
The last method, dynamic projection, involves updating each individual’s profile annually to reflect ageing and other changes. While longitudinal data at the individual-level would be ideal for capturing individual-level dynamics, such data is rarely available. Consequently, dynamic projection methods often rely on longitudinal data for one or a limited number of attributes to infer transition laws. In contrast, simulation models typically represent individuals through multiple interdependent attributes—for example, eight in \citet{horl_open_2021}. However, transition data for these attributes are generally scarce, which constrains dynamic projections to a reduced set of attributes—for instance, three in \citet{ballas_building_2005}, \citet{namazi-rad_generating_2014}, and \citet{kukic_hybrid_2023}.
To address this limitation, we propose a dynamic projection approach based on aggregated data, which is typically more readily available.
\redcomment{This approach is admittedly simplistic, as it does not capture complex interdependencies and is subject to stochastic variation. It is therefore not expected to achieve high efficiency. Its primary purpose is to enable a comprehensive comparison with existing methods and to assess the extent of performance achievable under such simplified conditions.}

%As a result, dynamic projection often relies on longitudinal data on one or a few attributes to derive transition laws.
%However, simulations typically model individuals using multiple interdependent attributes--e.g. eight in \citet{horl_open_2021}--yet. Transition data for these attributes is often scarce, resulting in dynamic projection with few attributes--e.g. three in \citet{ballas_building_2005, namazi-rad_generating_2014} and in \citet{kukic_hybrid_2023}.
%To address this limitation, we propose a dynamic projection method using aggregated data, which is more accessible. 
%\redcomment{This method is simplistic, as it does not account for complex relationships and is subject to randomness. Consequently, it is not expected to be highly efficient. Its purpose is to provide a comprehensive comparison with all existing methods, allowing us to assess what can be achieved.}

The quality of agent-forecasting also depends on the initial synthetic population. In population synthesis, several methods have been proposed to generate multi-dimensional individuals, using probabilistic models \citep{farooq_simulation_2013, sun_bayesian_2015}, or Deep Generative Models (DGMs) \citep{borysov_how_2019, garrido_prediction_2020, darsel_robust_2025}. An asset of these methods is their ability to create samples out of the training data, but also to generate an infinite pool of individuals, which can be useful for methods that copy individuals from the initial dataset. 
In this work, we investigate the impact of using a population generated by a Bayesian Network, a probabilistic model, and by a diffusion model, a DGM, on the forecasted population. 

Furthermore, most existing work on agent-forecasting represent individuals using only a limited set of attributes, despite the fact that simulations are capable of handling higher-dimensional representations, which can improve the simulation's quality. In this work, we examine how the different forecasting methods perform when the number of attributes increases. 
Moreover, previous research has typically evaluated generated populations on data from only one or two years, neglecting to investigate the evolution of the quality of the population over time. In this paper, we explore the performance of the different methods over multiple years.

This article addresses the key gaps identified in the existing literature by:
\begin{itemize}
    \item Assessing the reliance of synthetic populations from various sources (training data or generated by generative models) and evaluating their quality over multiple years.
    \item Proposing a method for dynamic projection when the transitions between states are not available, based on aggregated statistics.
    \item Evaluating the different methods (static projection, resampling, and dynamic projection) for agent-forecasting on multi-dimensional individuals over years.
    \item Comparing generative methods (Bayesian network, and diffusion model) for initializing synthetic populations and generating diverse sample pools in agent-forecasting.
\end{itemize}

In Section \ref{literature_review}, we present a literature review on agent-forecasting. The different projection methods are then derived in Section \ref{part:methodology}, with a closer attention to dynamic projection, for which we derive a new model. Then, in Section \ref{part:experiments} we describe our experiments framework, including the data used for the experiments, and the evaluation methodology. We present the results on evaluating population synthesis solutions over time and on comparing agent-forecasting methods. Some discussions about the results are given in Section \ref{part:discussions}.

% To that extent, we have access to the French census data, that are publicly available each year.

% Several models have been derived going from reproduction models \citep{beckman_creating_1996, voas_evaluation_2000, ye_methodology_2009} to probabilistic models \citep{farooq_simulation_2013, sun_bayesian_2015} and Deep Generative Models \citep{borysov_how_2019, garrido_prediction_2020, kim_deep_2023, darsel_population_2025}, allowing to generate samples out of the data, but also to model high-dimensional agents.

\section{Literature Review}
\label{literature_review}

Agent-forecasting methods can be categorized into three distinct approaches, discussed below. These approaches can be combined into hybrid models, which we address in the last subsection.

\subsection{Static projection}

Static models rely on a microdata sample of the population and adjust the weights of these samples to align with specific macroscopic parameters or dynamics for the targeted population. They generate populations directly from the initial dataset, without simulating intermediate steps or evolutionary processes.
\citet{ballas_geography_2005} explain that an advantage of this method is that it can allow to simulate "what-if" scenarios, thanks to the ability of static methods by imposing specific characteristics on the generated population.

%are re-weighting the population based on aggregate constraints, based on macro forecasts.

% This approach shares similarities with re-weighting methods in population synthesis, such as Iterative Proportional Filtering (IPF) \citep{beckman_creating_1996}, or Iterative Proportional Updating (IPU) \citep{ye_methodology_2009}. IPF approximates the joint distribution from a microsample and the marginal distributions, making it directly applicable to the current problem. The aim is to estimate the true proportion of each sample in the microsample. Starting from the initial proportions, the weights are iteratively adjusted to align with the marginal constraints. IPU extends this algorithm to sample households. Static projection applies these re-weighting methods to new constraints that differ from the original dataset.

\citet{ballas_modelling_2006} first generate a population with a population synthesis re-weighting technique, for the area of Leeds. Then, different scenarios on closing a factory in the area are tested, and modifications in the household attributes are directly applied. The impact on simulation is then analyzed with this modified population. Re-weighting is also performed by \citet{harding_demographic_2011}, who use the population projections in age, sex, and employment status together to re-weight the initial microsamples.
\citet{harding_demographic_2011} and \citet{tanton_spatial_2012} explain that re-weighting methods are only suitable for short term forecasts due to lack of long-term data that are mandatory to get accurate forecasts on the macro indicators. However, \citet{lomax_open-source_2022} explain how a re-weighting method can be efficient to test multiple long-term scenarios.  \citet{lomax_open-source_2022} adjust the weights based on the sex, age, and ethnicity of the population.

\subsection{Resampling}
Static projection can be computationally expensive, as it requires generating a new population for each year. To address this, \citet{predhumeau_synthetic_2023} propose to adapt it to a resampling method, which introduces some dynamic elements into the process. For the initial year, the population is reconstructed using the available microsample data. Subsequently, the population is updated annually by duplicating the population from the preceding year and then selectively copying or removing individuals to satisfy specific constraints. Unlike static projection, this method focuses on a subset of attributes rather than requiring all marginals. In \citet{predhumeau_synthetic_2023}, the joint distribution of age, location and gender is used. 

There are two main motivations for this method. Firstly, it avoids the computational cost of static projection that requires a model to be learnt for each year. Secondly, it can handle a larger number of attributes, as the population is updated based on a selection of attributes only -- 7 attributes in \citet{predhumeau_synthetic_2023}. However, it assumes that the correlations between individual attributes do not change much over time, relying on the initial pool of individuals for reconstruction. Thus, \citet{predhumeau_synthetic_2023} recommend this method only for short- and medium-term forecasts. By only copying individuals from the current population, there is a risk to decrease the diversity in the population, as some combinations of instances may disappear over time. 

\subsection{Dynamic projection}
Dynamic models start by sampling a population with a traditional population synthesis model. Then, each agent, either the household or the individual, is updated yearly by considering life events (birth, death, migration...).
This ageing process can be interpreted as an adaptation of the component-cohort method for the microscopic scale \citep{van_imhoff_microsimulation_1998}. Indeed, the component-cohort method models the evolution of a population at a macroscopic scale by considering demographic indicators (such as mortality or migration rates) for different sexes and age groups to update the total population at each time step. In dynamic agent-forecasting projection, this evolution is applied stochastically at the individual level, turning the rates into a probability. On average, the demographic indicators are respected.

Various demographic and socio-economic rules can be incorporated to update synthetic populations. For instance, \citet{van_imhoff_microsimulation_1998} account for mortality, fertility, and changes in marital status (union formation, separation, marriage, and divorce), using forecasts from Statistics Netherlands, although their model is restricted to a relatively small population of around 10,000 individuals. In contrast, \citet{vencatasawmy_building_1999} developed SVERIGE, a comprehensive model for the entire Swedish population, structured into ten modules that simulate diverse life events, including migration, education, and employment. However, SVERIGE lacks a systematic evaluation of its outputs. Similarly, \citet{ballas_building_2005} constructed a model for Ireland that integrates mortality, fertility, and internal migration. By explicitly modelling internal migration, their framework allows for the simulation of the entire national population, although it omits external migration flows. In this approach, internal migration is determined by the relative attractiveness of each county, which is assumed to be proportional to its population size.
More recent contributions, such as \citet{namazi-rad_generating_2014} and \citet{fatmi_baseline_2017}, adopt a two-step strategy. They first reconstruct the full population using population synthesis and re-weighting methods, and subsequently simulate its temporal evolution. Importantly, these studies model dynamics at both the household and individual levels, thereby offering a more detailed representation of population change.

\subsection{Hybrid approaches}
Although the resampling may look attractive, it requires knowledge of the joint distribution, which may not be available annually. Therefore, \citet{kukic_hybrid_2023} propose a hybrid approach, between resampling and dynamic projection. The method begins with an initial population, which is then updated using resampling when joint distribution data is available and dynamic projection otherwise. Unlike \citet{predhumeau_synthetic_2023}, this approach first ages the population by incrementing each individual’s age by one year, before applying resampling.

When an additional microsample becomes available, the resampling technique can be adapted by incorporating samples from the new microsample instead of the initial dataset \citep{kukic_adaptive_2025}. In their work, \citet{kukic_adaptive_2025} introduced several refinements to the resampling process. They employed a Monte Carlo Markov Chain sampler as proposed by \citet{kukic_one-step_2024}, and they used the household size as a criterion for resampling. Additionally, they generated new individuals from the new microsample to match the desired marginal distributions. Excess samples are then removed to maintain the population size while preserving the required proportions. Since resampling allows to handle a higher number of attributes, \citet{kukic_adaptive_2025} increase the number of attributes simulated to 8. 
To the best of our knowledge, in the literature, \citet{kukic_adaptive_2025} is the only model that deals with microsamples from several years--two years are used in the training and one year for the evaluation, and it is also the only model with eight attributes which corresponds to the number of attributes generated by eqasim \citep{horl_synthetic_2021}.

\section{Methodology}
\label{part:methodology}

Building on the work of \citet{kukic_adaptive_2025}, we aim to compare the ability of various methods to generate populations ready for simulation, i.e., with a sufficient number of attributes. Existing literature on this topic typically relies on datasets with limited temporal coverage, often using only a single year of data for evaluation.
In this paper, we assess the ability of different projection methods to forecast population over multiple years. Our approach uses microsample from only one year, combined with aggregated statistics from subsequent years. 

In this section, we describe the derivation of the different models for agent-forecasting. To that extent, we first introduce some notations that are similar for all models. Then, we present the attributes that we want to simulate in the models\redcomment{, as some of them appear in the derivation of the methods}.

\subsection{Notations}
To derive mathematically the different models and the evaluation metrics, we introduce a couple of notations.
Let ${}^y\mathbf{X}= \left({}^yX_1, {}^yX_2, \dots, {}^yX_d\right)$ denote the $d$-dimensional distribution of the population for year $y$, which we aim to approximate. A set of $n$ realizations of ${}^y\mathbf{X}$ is represented by ${}^y\mathbf{S}^n$, where each sample ${}^y\mathbf{s}^i = \left({}^y{s^i}_1, {}^y{s^i}_2, \dots, {}^y{s^i}_d\right)$ (for $i \in [1, n]$) corresponds to an individual realization. We denote ${}^y\tilde{\mathbf{S}}$ as a set  of samples approximating  a set of realisations from ${}^y\mathbf{X}$. In static projection and resampling, the size of the the generated set is controllable, and we note $N_y=|{}^y\tilde{\mathbf{S}}|$ the target size of this set. For $k\in[1,d]$, $\Omega_k$ corresponds to all modalities that the attribute ${}^{y}X_k$ could take over all possible years. This set is independent from the year. For a given modality $\omega_k \in \Omega_k$, $f_k(\mathbf{S}, \omega_k)$ denotes the frequency of $\omega_k$ in the set $\mathbf{S}$.
For clarity, we re-index the years such that the initial year, i.e. the year where data is available, is set to $y=0$.

\subsection{Attributes}

In our experiments, we consider eight attributes to describe an individual (three numerical and five categorical attributes). These attributes are presented in Table \ref{tab:attributes_description}, and are in line with simulation \citep{horl_synthetic_2021}. This set is bigger than most of the literature on agent-forecasting, but it is consistent with common practice in population synthesis, where handling eight or more attributes is standard.
% ompared to the literature, this amount of attributes is higher than most agent-forecasting experiments, but it corresponds to a median number for population synthesis, and corresponds to some simulation setups \citep{horl_synthetic_2021}.

\begin{table}[t]
\scriptsize{
    \renewcommand*{\arraystretch}{1.5}
    \centering
    % \begin{tabular}{|p{5cm}|l|p{1.7cm}|c|c|}
    \begin{tabular}{|l|l|c|c|c|}
    \hline
     Name & Type & Number of instances & min & max\\
    \hline
    Age & integer & 120 & 0 & 120 \\
    Sex & binary & 2 &  & \\
    Last diploma & category & 5 &  & \\
    Number of persons in the household & integer & 25 & 0 & 64 \\
    % Type of household & category & 5 &  & \\
    Type of professional activity & category & 9 &  &  \\
    % Family link & category & 5 &  & \\
    Married & boolean & 2 &  & \\
    Department & category & 8 &  &  \\
    Number of cars & integer & 4 & 0 & 3  \\
    % Socio-professional category & category & 8 & &  \\
    % Type of accommodation & category & 7 &  & \\
    \hline
    \end{tabular}
    \caption{Description of the studied attributes}
    \label{tab:attributes_description}
    }
\end{table}

\subsection{Derivation of static projection (idealistic)}
\label{part:static_ideal}
Static projection requires access to an initial set of realizations from the first year, denoted by ${}^0\mathbf{S}^n$, and some margins for the desired year:
\[
\forall k \in[1,d], \forall\omega_k\in\Omega_k, \mathbbm{P}({}^y{X}_k = \omega_k)
\]

The generated population consists of samples drawn from the microsample, with each sample’s occurrence calculated to respect the specified marginal distributions. Mathematically, for a given year $y$, the generated population can be expressed as:
\begin{equation}
    \begin{split}
        &{}^y\tilde{\mathbf{S}} = \bigcup_{i=1}^n \left( {}^0\mathbf{s}^i \text{ repeated } {}^y{w}^i \text{ times} \right) \\
        &\text{such that }
        \begin{cases}
            \forall k \in[1,d], \forall\omega_k\in\Omega_k, f_k\left({}^y\tilde{\mathbf{S}},\omega_k\right)=\mathbbm{P}\left({}^y{X}_k = \omega_k\right) \\
            |{}^y\tilde{\mathbf{S}}| = \sum_{i=1}^n {}^y{w}^i = N_y
        \end{cases}
    \end{split}
\end{equation}
where ${}^y{w}^i \in \mathbbm{N}$ corresponds to the occurrence of ${}^0\mathbf{s}^i$ in ${}^y\tilde{\mathbf{S}}$. The constraints specifies the margins and the size of the desired population.

To integrate the margins to a microsample, Iterative Proportional Fitting (IPF) is a common method in population synthesis \citep{beckman_creating_1996, ballas_building_2005, ye_methodology_2009}. IPF approximates the joint distribution from a microsample and the marginal distributions, making it directly applicable to the current problem. The aim is to estimate the true proportion of each sample in the microsample. Starting from the initial proportions, the weights are iteratively adjusted to align with the marginal constraints. This operation is performed until an error termination criterion is reached.

To improve the quality of the microsample and ensure precision, additional bins may be introduced for specific attributes during the IPF procedure. This adjustment mitigates potential bias caused by an insufficient number of samples in some instances. For example, in our experiments, we create age bins because the microsample contains very few individuals over 100 years old. Without these bins, some age groups might be underrepresented or missing from the microsample, even though they are relevant for certain years. To address this, we used 15-year age intervals.

To obtain the population with the desired marginals, the individuals are picked to fit the estimated proportion. When the desired count $c$ is not an integer, stochastic rounding is applied \citep{gupta_deep_2015}.

\redcomment{The described approach is idealized, as it assumes full knowledge of all marginal distributions—an impractical scenario in real-world applications. In practice, projections for these margins are often inaccurate and unavailable for many attributes. Nevertheless, we remain interested in assessing the potential performance of a method operating under such ideal conditions.}

% % ¨Pas forcément besoin du code ici

% \begin{algorithm}
% \caption{Static Projection algorithm}\label{alg:static_proj}
% \begin{algorithmic}
% \Require ${}^0\mathbf{S}^n$; ${}^yN$; $e_{max}$; $\forall k \in[1,d], \forall\omega_k\in\Omega_k, \mathbbm{P}({}^y{X}_k = \omega_k)$
% \State ${}^y\tilde{\mathbf{S}} = []$
% \For{$i \gets 1$ to ${}^yN$}
%     \State $\bar{k} = \argmax_k SRMSE_k\left({}^y\tilde{\mathbf{S}}, {}^y{X}_k\right) $
%     \State $\bar{\omega} = \argmin_{\omega_{\bar{k}} \in \Omega_{\bar{k}}} \left(f_{\bar{k}}\left({}^y\tilde{\mathbf{S}},(\omega_{\bar{k}})\right) - \mathbbm{P}\left({}^y\mathbf{X}_{\bar{k}} = \omega_{\bar{k}}\right)\right)$

%     \State $\mathbf{s} = pick\_sample\_with\_modality\left({}^0\mathbf{S}^n, \bar{k}, \bar{\omega}\right)$
%     \Comment{Pick randomly a sample $\mathbf{s}$ from ${}^0\mathbf{S}^n$, such that $s_{\bar{k}} = \bar{\omega}$}
    
%     \State ${}^y\tilde{\mathbf{S}} \mathrel{+}= \mathbf{s}$ \Comment{Add the selected sample to the resulting population}
% \EndFor
% \State \Return ${}^y\tilde{\mathbf{S}}$
% \end{algorithmic}
% \end{algorithm}

\subsection{\redcomment{Derivation of static projection (realistic)}}
% \redcomment{The method presented in Section \ref{part:static_ideal} is idealistic. It supposes that we have access to the true values of the margins for all attributes, for the future years. This is not realistic in practice, as the projections cannot be perfectly accurate, and are usually done only on some attributes, but not all attributes.}

\redcomment{Since the method described in Section \ref{part:static_ideal} lacks realism, we propose a more practical alternative based on the same underlying principle. Introducing artificial degradation to true values would make the results excessively sensitive to experimental conditions, so we avoid altering prediction accuracy. Instead, we only adjust the number of attributes considered.}

\redcomment{In this approach, we fit the IPF using all attributes for which projections are available from the French National Institute of Statistics and Economic Studies (INSEE): sex, age, and department \footnote{\href{https://www.insee.fr/fr/outil-interactif/6798992/pyramide.htm}{Age Pyramid - INSEE}}. While this framework remains optimistic, by assuming access to true values, it is designed for practical implementation.}

\subsection{Derivation of the resampling approach}

The resampling approach requires access to an initial solution ${}^0\mathbf{S}^n$, and some yearly constraints. In this paper, we will consider the same constraints as in \citet{predhumeau_synthetic_2023}, by resampling with the true joint distribution for the age, the sex and the location (${}^yX_{age,sex,loc}$). But unlike \citet{predhumeau_synthetic_2023}, we do not copy samples from the current population, but from an initial wide pool of individuals. This allows to solve the potential risk of vanishing some profiles, and keep diversity over time.

The initial solution can be obtained with any algorithm from population synthesis ${}^0\tilde{\mathbf{S}} = g\left({}^0\mathbf{S}^n\right)$, where $g$ is the function associated to the algorithm, which enables to construct the full population from a microsample.
Then for a new year $y$, we copy the population from the previous year: ${}^y\bar{\mathbf{S}} = {}^{y-1}\tilde{\mathbf{S}}$. 
% We crop the population in order to remove surplus items in each category:
% \begin{equation}
%     \begin{split}
%     &{}^y\bar{\bar{\mathbf{S}}} = \argmax_{\mathbf{T}} |\mathbf{T}_1|\\
%     \text{ such that }&\begin{cases}
%         \mathbf{T}\subseteq {}^y\bar{\mathbf{S}}\\
%         \forall \omega_{age,sex,loc} \in \Omega_{age,sex,loc}, \mathbbm{P}\left({}^yX_{age,sex,loc} = w_{age,sex,loc}\right)\geq f_{age,sex,loc}\left(\mathbf{T},\omega_{age,sex,loc}\right)
%     \end{cases}
%     \end{split}
% \end{equation}
% And we add some samples from the \textbf{initial microsample}, to have the desired proportion:
% \begin{equation}
%     \begin{split}
%     &{}^y\tilde{{\mathbf{S}}} = {}^y\bar{\bar{\mathbf{S}}} \cup \mathbf{T}_2\\
%     \text{ such that }&\begin{cases}
%         \forall \mathbf{s} \in \mathbf{T}_2, \mathbf{s} \in {}^0{\mathbf{S}}^n\\
%         \forall \omega_{age,sex,loc} \in \Omega_{age,sex,loc}, \mathbbm{P}\left({}^yX_{age,sex,loc} = w_{age,sex,loc}\right) = f_{age,sex,loc}\left({}^y\tilde{{\mathbf{S}}},\omega_{age,sex,loc}\right)\\
%         |{}^y\tilde{{\mathbf{S}}} | = N_y
%     \end{cases}
%     \end{split}
% \end{equation}
% In practice, the last two steps can be performed together. 
For each combination of age, sex and location, the desired number of samples in the population is computed, and compared with the current count of samples with that combination. If the current count exceeds the desired number, some samples are randomly removed from the population. Otherwise, if the number of samples is below the desired number, samples are randomly drawn from the initial microsample. 
%The implementation is described in Algorithm \ref{alg:resampling}.

We note that the processes are independent for each group of age, sex and location. Therefore, if we want to sample to one specific year, we can directly sample to the target year from year $0$, with the same process.

As for static projection, this refitting algorithm could require using bins. We use the same bins as for the static projection.

\redcomment{While this method remains idealistic in assuming access to true values, it is realistic to assume access to INSEE’s projections of the joint distribution for age, sex, and department\footnote{\href{https://www.insee.fr/fr/outil-interactif/6798992/pyramide.htm}{Age Pyramid - INSEE}}. These projections are the same data used in our realistic static projection method, but here we use the joint distribution for resampling—unlike the static method, which only considers the marginal distributions.}

% \begin{algorithm}
% \caption{Resampling algorithm}\label{alg:static_proj}
% \begin{algorithmic}
% \Require ${}^0{\mathbf{S}}^n$;${}^0\tilde{\mathbf{S}}$; ${}^yN$; $\forall\omega_{age,sex,loc}\in\Omega_{age,sex,loc}, \mathbbm{P}({}^y\mathbf{X}_{age,sex,loc} = \omega_{age,sex,loc})$
% \State ${}^y\bar{\mathbf{S}} = {}^{0}\tilde{\mathbf{S}}$
% \For{$\omega_{age,sex,loc}$ in $\Omega_{age,sex,loc}$}
%     \State target = $\mathbbm{P}\left({}^yX_{age,sex,loc} = \omega_{age,sex,loc})\right) \times N_y$
%     \Comment{}
%     \State current\_number = $\left|\{{}^{y-1}\mathbf{s}; {}^{y-1}{s}_{age,sex,loc}=\omega_{age,sex,loc}\}\right|$ 
%     \If{current\_number$>$target}
%         \State ${}^y\bar{\mathbf{S}} = remove\_samples({}^y\bar{\mathbf{S}},\omega_{age,sex,loc},$current\_number$-$target$)$
%     \EndIf
%     \If{current\_number$<$target}
%         \State ${}^y\bar{\mathbf{S}} = add\_samples({}^y\bar{\mathbf{S}},{}^0{\mathbf{S}},\omega_{age,sex,loc},$target$-$current\_number$)$
%     \EndIf
%     \State ${}^{y}\tilde{\mathbf{S}} = {}^{y}\bar{\mathbf{S}}$
% \EndFor
% \State \Return ${}^y\tilde{\mathbf{S}}$
% \end{algorithmic}
% \label{alg:resampling}
% \end{algorithm}

\subsection{Derivation of dynamic projection}

In the general case, dynamic projection requires access to an initial solution ${}^0\mathbf{S}^n$, and some updating rules ${}^yu_1,...{}^yu_r$ for each year. 
As for the resampling, a population is reconstructed for the initial year ${}^0\tilde{\mathbf{S}} = g\left({}^0\mathbf{S}^n\right)$. %, where $g$ is the function associated to any algorithm from population synthesis, which enables to construct the full population from a microsample.
From this initial population, individuals are aged each year. To construct the population for the year after ${}^{y+1}\mathbf{S}^i$, each updating rule is applied one after the other on each individual to get the individual for the next year:
\[
    {}^{y+1}\mathbf{s}^i={}^yu_r\circ...\circ{}^yu_1({}^{y}\mathbf{s}^i).
\]
The individual is kept in the population, based on mortality and migration rules. In our model, we omit the latter, because of the lack of data. We only consider internal migration.

In our experiments, we do not benefit from longitudinal data for the full joint distribution, but we have a representation of the population for each year. In order to simulate a dynamic scheme, we construct heuristic rules fitting the aggregated statistics. While this approach approximates a dynamic model, it may be methodologically less robust than traditional dynamic projection methods relying on longitudinal data.
We distinguish four types of attributes:
\begin{itemize}
    \item attributes with a direct relation with the year and the previous situation, e.g. age. This first class of attributes does not require any detailed explanations, as the derivation of the updating rule is straightforward.
    \item attributes that can be computed from external data sources, e.g. marriage rate. The rates are often given for each age. Therefore, we derive the updating rules conditionally to the age.
    \item attributes constrained by a chronological order between the instances, e.g. highest diploma.
    \item attributes with no or few constraints, e.g. professional activity or the department.
\end{itemize}

\begin{figure}
    \centering
    \includesvg[width=0.9\linewidth]{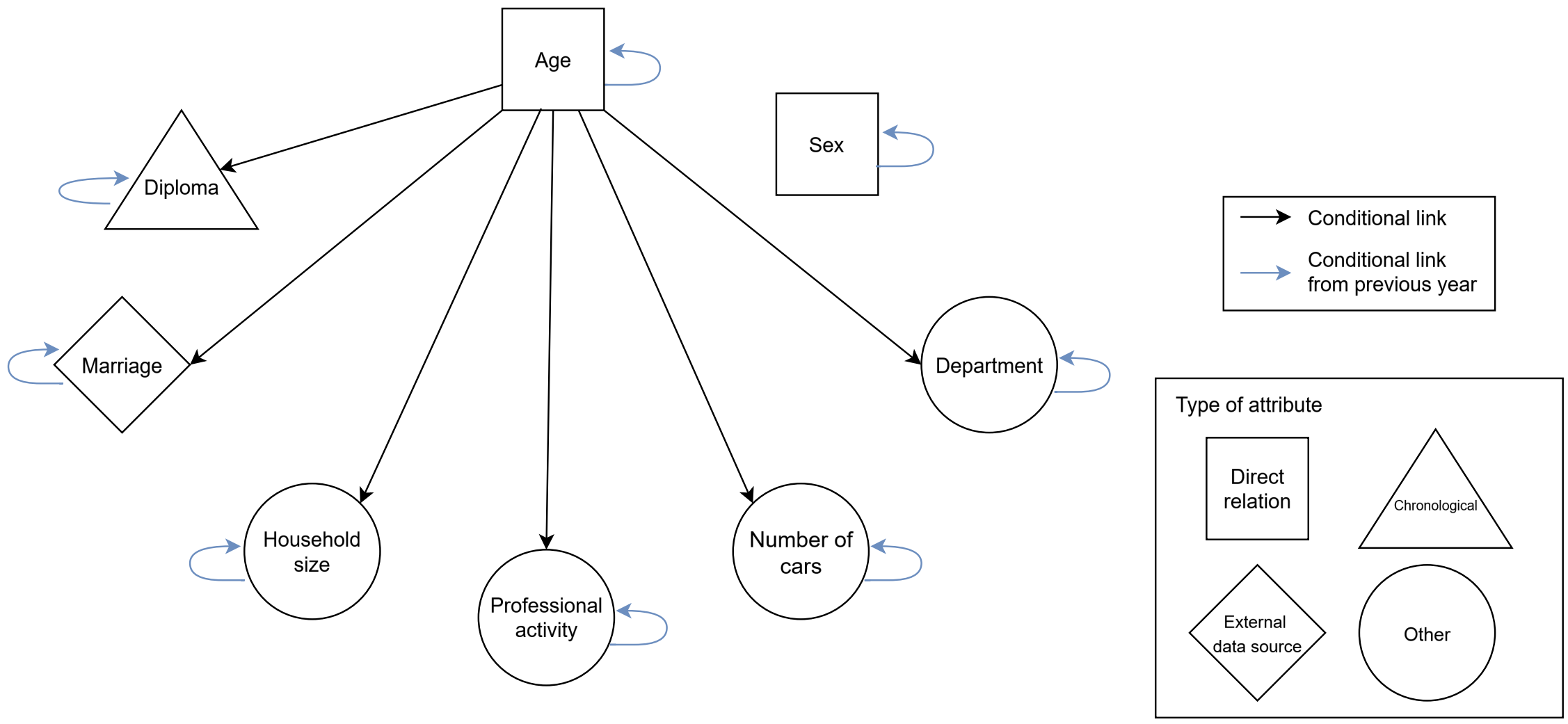}
    \caption{Attributes used in our experiments with their corresponding type, and the conditional links used for the derivation}
    \label{fig:attributes_description_dynamic}
\end{figure}

The latter two require more attention because of the lack of longitudinal or additional data. Figure \ref{fig:attributes_description_dynamic} presents the categorization of attributes used in our experiments, along with the conditional dependencies applied.
In our model, each attribute depends on its previous state. Additionally, all attributes except the sex are conditioned by age, reflecting its central role in the individual development.

\subsubsection{Updating attributes with ordered modalities}
If there is an order between the modalities, no return to a previous state is possible.
Therefore, the transition matrix is triangular, reducing the number of unknown quantities.

Among ordered attributes, some have sequential modalities, meaning an individual can only progress to the next modality in a fixed order. Age is an example of attribute that follows this rule: its modalities are  strictly ordered, and an individual can only transition from one modality to the immediately subsequent one (or remain in the same modality). For such attributes, the transition matrix has non-null values only on the main diagonal and on the first diagonal upper the main diagonal. In the case of the age, there is no need to compute the transition matrix, as our temporal step is one year, so the transitions are straightforward.

In our experiments, we only have one other attribute that have ordered modalities: the highest degree. For this attribute, each modality can be reached only from the previous modality, with an exception. The highest degree can be obtained from two possible states: holding a high school diploma or a professional diploma. Therefore, in the transition matrix, we have one element non-null outside the two principal diagonals of the triangular matrix.

The different transition matrix corresponding to the different cases mentioned above are given here:

\begin{multicols}{3}
Transition matrix for an ordered attribute:
% \begin{equation*}
%     \begin{pmatrix}
%         m_{1,1} & m_{1,2} & \cdots & m_{1,n-1} & m_{1,n}\\
%         0 & m_{2,2} & \cdots & m_{2,n-1} & m_{2,n} \\
%         \vdots & \ddots & \ddots & \ddots & \vdots \\
%         % \vdots & \vdots & \ddots & \vdots \\
%         \vdots & \ddots &\ddots & m_{n-1,n-1} & m_{n-1,n} \\
%         0 & \cdots &\cdots & 0 & m_{n,n}
%     \end{pmatrix}
% \end{equation*}
\begin{equation*}
    \begin{pmatrix}
        m_{1,1}  & \cdots &\cdots & m_{1,n}\\
        0 & \ddots &  & \vdots \\
        \vdots & \ddots & \ddots & \vdots \\
        % \vdots & \vdots & \ddots & \vdots \\
        % \vdots & \ddots &\ddots & m_{n-1,n-1} & m_{n-1,n} \\
        0 & \cdots & 0 & m_{n,n}
    \end{pmatrix}
\end{equation*}
\columnbreak

Transition matrix for an ordered attribute, when the modalities are mandatorily successive:
\begin{equation*}
    \begin{pmatrix}
        m_{1,1}  & m_{1,2} & 0 &\cdots & 0\\
        0 & \ddots & \ddots & \ddots & \vdots \\
        % \vdots & \ddots & \ddots & \ddots & \vdots \\
        % \vdots & \vdots & \ddots & \vdots \\
        % \vdots & \ddots &\ddots & m_{n-1,n-1} & m_{n-1,n} \\
        \vdots & \ddots & \ddots & \ddots & 0 \\
        \vdots &  & \ddots & m_{n-1,n-1} & m_{n-1,n} \\
        0 & \cdots & \cdots & 0 & m_{n,n}
    \end{pmatrix}
\end{equation*}
\columnbreak

Transition matrix for our degree attribute (successive modalities with one exception):
\begin{equation*}
    \begin{pmatrix}
        m_{1,1}  & m_{1,2} & 0 &\cdots &\cdots & 0\\
        0 & \ddots & \ddots & \ddots & m_{i,j} &\vdots \\
        % \vdots & \ddots & \ddots & \ddots & \vdots \\
        % \vdots & \vdots & \ddots & \vdots \\
        % \vdots & \ddots &\ddots & m_{n-1,n-1} & m_{n-1,n} \\
        \vdots & \ddots & \ddots & \ddots &\ddots &  \vdots\\
        \vdots &  & \ddots & \ddots &\ddots &  0 \\
        \vdots&  &  & \ddots & m_{n-1,n-1} & m_{n-1,n} \\
        0 & \cdots& \cdots & \cdots & 0 & m_{n,n}
    \end{pmatrix}
\end{equation*}
\end{multicols}

To compute the transition probabilities for the highest diploma, we use the observed frequencies of age and diploma combinations for two consecutive years. We can deduce the transition coefficients to fit the margins. However, we have one degree of freedom in our equation systems due to the extra coefficient.  We assume that individuals with a high school diploma are prioritized for admission to higher education. If the target margin for higher education diplomas is not met by individuals who already hold such a diploma or those with a high school diploma, the remaining share is fulfilled by individuals holding a professional diploma.

% have degree of freedom in the equation  a given a

% according to the order, from the previous instance in order to fit the marginal. Once the last instance is fitted, we can do the same process recursively.ttribute, we use the joint distribution from this attribute and the age for two consecutive years. 

% In our experiments, only the diploma is concern by this ...; . The highest degree can be obtained from two possible states: holding a high school diploma or a professional diploma, both of which can lead to the final state of obtaining a higher education diploma. We assume that individuals with a high school diploma are prioritized for admission to higher education. If the target margin for higher education diplomas is not met by individuals who already hold such a diploma or those with a high school diploma, the remaining share is fulfilled by individuals holding a professional diploma.

\subsubsection{Updating attributes with unordered modalities}
For the last category of attributes, we also use the joint distribution between the attribute of interest and the age for two consecutive years. This time, there are too many transition coefficients to compute. The equation system has no unique solution. So we derive an heuristic method based on two main motivations: respecting the distribution, and preserving a momentum for individual states. The momentum represents that individuals tend to remain in the same state over years, so they are likely to remain with the same characteristics.

To that extent, we initialize the transition for each attribute ${}^yX_k$ to remain in state $\omega_k$ using a momentum rate $m_k\in[0,1]$. The remaining probability mass is then distributed among all possible states (including the previous one) to align with the marginal distributions.
In practice, we adapt this momentum rate, for each state, in order to avoid exceeding the margin:
\begin{equation}
    M_{\omega_k} = \min\left(m_k,\frac{\mathbbm{P}({}^{y+1}{X}_k=\omega_k)}{\mathbbm{P}({}^{y}{X}_k=\omega_k)}\right).
    \label{eq:momentum}
\end{equation}
Thus, the transition probability is given by:
\begin{equation}
    \mathbbm{P}\left({}^{y+1}{X}_k=\omega'_k|{}^{y}{X}_k=\omega_{k}\right) = M_{\omega_k}\mathbbm{1}_{\omega'_k=\omega_k} + (1-M_{\omega_k})\frac{\mathbbm{P}({}^{y+1}{X}_k=\omega'_k)-M_{\omega'_k}\times\mathbbm{P}({}^{y}{X}_k=\omega'_k)}{\sum_{\omega\in\Omega_k}\mathbbm{P}({}^{y+1}{X}_k=\omega)-M_{\omega}\times\mathbbm{P}({}^{y}{X}_k=\omega)}
    \label{eq:pot}
\end{equation}
% A choisir entre les deux
% \begin{equation}
%     \mathbbm{P}\left({}^{y+1}{X}_k=\omega'_k|{}^{y}{X}_k=\omega_{k}\right) = M_{\omega_k}\mathbbm{1}_{\omega'_k=\omega_k} + (1-M_{\omega_k})\frac{\mathbbm{P}({}^{y+1}{X}_k=\omega'_k)-M_k\times\mathbbm{P}({}^{y}{X}_k=\omega'_k)}{1-\sum_{\omega\in\Omega_k}M_k\times\mathbbm{P}({}^{y}{X}_k=\omega)}
%     \label{eq:pot2}
% \end{equation}
In Equation \ref{eq:pot}, the first term corresponds to the momentum term and the second term to the random assignment of the remaining weights, so the marginal distributions are respected.

In our experiments, as shown in Figure \ref{fig:attributes_description_dynamic}, we condition Equation \ref{eq:pot} on the individual’s age, as an individual’s characteristics—-such as student status—-are highly age-dependent. To better capture the evolving dynamics of each individual and keep realist individuals, the marginal distributions are therefore conditioned on age.

\subsubsection{Parameters for not ordered attributes}

\begin{table}[t]
\scriptsize{
    \renewcommand*{\arraystretch}{1.5}
    \centering
    % \begin{tabular}{|p{5cm}|l|p{1.7cm}|c|c|}
    \begin{tabular}{|l|l|c|}
        \hline
         Name & Momentum  \\
        \hline
        % Age & Direct & \\
        % Sex & Direct& \\
        % Last diploma & Chronological& \\
        Number of persons in the household &0.99\\
        % Type of household & Not chronological& 0.99\\
        Type of professional activity & 0.95\\
        % Family link & Not chronological&0.99\\
        % Married & External &\\
        Department & 0.98\\
        Number of cars & 0.9\\
        % Socio-professional category & Not chronological&0.95\\
        % Type of accommodation & Not chronological&0.95\\
        \hline
        \end{tabular}
    \caption{Momentum values chosen for the attributes with no order between the instances.}
    \label{tab:attributes_description_dynamic}
    }
\end{table}

Unlike other the attribute types, each non-ordered attribute requires a hyperparameter to define its momentum value. We have chosen the values based on the expected stability of the instances. The chosen momentum values for each of the 4 non-ordered attributes are provided in Table \ref{tab:attributes_description_dynamic}.

By setting the momentum parameter to values below 1, we introduce flexibility that allows for exploration of alternative values and spontaneous transitions between states. The momentum rates are intentionally set high to maintain strong continuity between successive states. Additionally, in cases where there is a significant change in the margin of a given instance, the momentum coefficient is capped by the margin for the following year in Equation \ref{eq:momentum}. This ensures that redistribution remains possible when the share of an instance decreases.

~

Finally, we can derive an updating function for any of the four types of attributes. \redcomment{Since the last two categories of attributes are largely random and no cross-correlation—except with age—is considered, this method should effectively capture the dynamics of each attribute individually, but may struggle to represent their combined interactions.}

\subsection{Generation of the initial solution}
\label{part:meth_initial_sol}
All different models require access to a microsample ${}^0\mathbf{S}^n$. In static projection and resampling, the samples are replicated to generate the new populations, considering some constraints. However, the number of samples in the initial microsample is limited. Therefore, the generated populations could lack diversity, as it is impossible to produce samples outside the scope of the training data. This problem is known as the zero-cell problem in the population synthesis literature \citep{guo_population_2007}.

The extensive literature on population synthesis shows that there are reliable methods that enable a a good population reconstruction, thus preventing a lack of diversity. In this work, we propose changing the initial microsample ${}^0\mathbf{S}^n$ to another set drawn from ${}^y\mathbf{X}$ using state-of-the-art population synthesis algorithms. Based on \citet{darsel_population_2025}, we propose to use two different models--Bayesian Network and a Diffusion Model--to generate the initial microsample ${}^0\mathbf{S}^n$.
%${}^0\mathbf{S}^n$. ${}^0\mathbf{S}^N$, where  $N>>n$, in addition to keeping ${}^0\mathbf{S}^n$. 
These generative models generate an unlimited number of individuals. We leverage this ability to adapt the size of the initial microsample, but also to create a large pool of synthetic individuals, thereby enhancing the diversity and range of choices available in static projection and resampling models. This corresponds to generate ${}^0\mathbf{S}^N$, where  $N>>N_y$, with $N_y$ target size of the population for year $y$.
% Therefore, we can even oversample. 
% For the initial microsample, it allows to change the number of samples $n$. Moreover, we can generate a sample reserve with many more samples than the desired size of the population. This allows avoid having a sample selected multiple times in static projection and resampling. 
% The two models--Bayesian Network and a Diffusion Model--are also used to generate the initial solution in dynamic projection and resampling. 
% Moreover, these methods can generate an unlimited number of individuals. We leverage this ability to create a large pool of synthetic individuals, thereby enhancing the diversity and range of choices available in static projection and resampling models.

In static projection, changing the initial microsample directly impacts the results, as the initial solution in the IPF algorithm is derived from the proportion observed in the microsample. 

In our experiments, we use both the samples generated by generative models, and samples from the training data as the starting solution and the pool of individuals. This allows us to assess the impact of each method.

Details on the parameters of the generative models are given in \ref{appendix:pop_synth}.

\section{Results}
\label{part:experiments}
\subsection{Data}

In our experiments, we use open-source French census data, which is published every year. This disaggregated data comprises the entire population. The attributes describing the individuals are given in Table \ref{tab:attributes_description}. They correspond to the more relevant attributes available for an agent-based simulation model, supported by \citet{horl_open_2021}. In our experiments, we use the data from 2013 to 2021 provided by INSEE\footnote{The 2013 dataset: \url{https://www.insee.fr/fr/statistiques/2409376}. The other years are available on the website with the same name expect the year.}. We concentrate on the Île-de-France area, which is the most densely populated region in France.

When the population is projected, only a microsample from 2013 is used for training generative models, or directly as initial solution. Data from 2013 to 2021 is used for evaluation, but also to train reference models to ensure they continue to accurately represent the true population over time. Finally, data from 2014 to 2021 is used to compute the aggregated statistics required for implementing the different projection models and for evaluation purposes. We consider an optimistic scenario in which all these statistics are available and accurate, which is not necessarily the case in practice.

To fit with the literature on population synthesis, we restrict the available microsamples to 1\% of the total population. The 1\% dataset is constructed with stratified sampling, based on the age, the sex and the department, allowing to keep the true joint distribution for these attributes. In experiments involving a generative model, as presented in Section \ref{part:meth_initial_sol}, we are free in the size of the population we generate, as the generation of an individual is independent from the others. For the initial solution, we generate a population representing 5\% of the total population, to achieve a better and more diverse representation of the population. A larger population would increase the evaluation time.

In addition to the initial solution for 2013, we use generative models to create a pool of potential individuals that can be selected using methods that replicate individuals. We set this pool at 50\% of the total population size to ensure variety in our simulations. Although we ultimately generate a forecasted population of 5\%, maintaining a larger pool helps to prevent the same individuals from being selected multiple times.

Finally, for the dynamic projection method, some external data sources are used to simulate the birth rates\footnote{\url{https://www.insee.fr/fr/statistiques/7673408?sommaire=7673431}}, the mortality rates \footnote{\url{https://www.insee.fr/fr/statistiques/series/103039135}}, the marriage rates \footnote{\url{https://www.insee.fr/fr/statistiques/8570890?sommaire=8571148}} and the divorce rates \footnote{Data from 2013 only: \url{https://www.insee.fr/fr/statistiques/fichier/7624542/fm_t28.xlsx}}. All these external data sources are also provided by INSEE. We note that the divorce rates data is no longer published yearly, therefore we consider the figures from 2013 for every year of the simulation. Using the dynamic of one year of data over multiple years is commonly done in dynamic projection \citep{ballas_building_2005, kukic_adaptive_2025}.

\subsection{Evaluation}
To evaluate a generated population from a given year, we use a test set comprising 90\% of the total population, which is an extremely accurate representation of the whole population. 
We follow the evaluation criteria from \citet{darsel_robust_2025}. Privacy is not a concern in our case, as comparing with training data from another year is meaningless in forecasting. However, we consider the other two criteria: distribution and realism.

%To evaluate a generated population of a given year, we use a testing set composed of 90\% of the total population, what is a very good approximation of the total population. 
%We follow the evaluation criteria from \citet{darsel_robust_2025}. The privacy is out of the concern in our case, as in forecasting, comparing with the training data from another year, is meaningless. However, we consider the two other criteria: distribution and realism.

To evaluate the distribution, we consider the Standardized Root Mean Squared Error ($SRMSE$) on the marginals, the bi-variate and the tri-variate distributions. We consider the same metric as \citet{lederrey_datgan_2022} and \citet{darsel_robust_2025} by taking the average score on all combinations of attributes for each size of the $SRMSE$.

First, we introduce the $SRMSE$ for a group of one, two or three variables $X_k$, $X_l$, and $X_m$ between the generated data $\mathbf{X}_{gen}$ and the testing data $\mathbf{X}_{test}$, with:
\begin{equation}
    \begin{split}
        {SRMSE_{k}(\mathbf{X}_{gen},\mathbf{X}_{test}) = } &{\sqrt{ \sum\limits_{\substack{
        \omega_{k}\in \Omega_{k}}}
   \left(f_{gen}(\omega_{k})-f_{test}(\omega_{k})\right)^2\times\mid\Omega_k\mid}},\\
        {SRMSE_{kl}(\mathbf{X}_{gen},\mathbf{X}_{test}) = } &{\sqrt{ \sum\limits_{\substack{
                \omega_{kl}\in \Omega_{kl}}}
   \left(f_{gen}(\omega_{kl})-f_{test}(\omega_{kl})\right)^2\times\mid\Omega_k\mid\times\mid\Omega_l\mid}},\\
        {SRMSE_{klm}(\mathbf{X}_{gen},\mathbf{X}_{test}) =  }&{\sqrt{ \sum\limits_{\substack{
        \omega_{klm}\in \Omega_{klm}}}
   \left(f_{gen}(\omega_{klm})-f_{test}(\omega_{klm})\right)^2\times\mid\Omega_k\mid\times\mid\Omega_l\mid\times\mid\Omega_m\mid}}.        
    \end{split}
\end{equation} 
Thus, we can define the average score:
\begin{equation}
    \begin{split}
        \overline{SRMSE}_1(\mathbf{X}_{gen},\mathbf{X}_{test})=&\frac{1}{d}\sum\limits_{k=1}^dSRMSE_{k}(\mathbf{X}_{gen},\mathbf{X}_{test}), \\
        \overline{SRMSE}_2(\mathbf{X}_{gen},\mathbf{X}_{test})=&\frac{1}{\binom{d}{2}}\sum\limits_{(k,l)\in\binom{\{1,...,d\}}{2}}SRMSE_{kl}(\mathbf{X}_{gen},\mathbf{X}_{test}),\\
        \overline{SRMSE}_3(\mathbf{X}_{gen},\mathbf{X}_{test})=&\frac{1}{\binom{d}{3}}\sum\limits_{(k,l,m)\in\binom{\{1,...,d\}}{3}}SRMSE_{klm}(\mathbf{X}_{gen},\mathbf{X}_{test}).\\ 
    \end{split}
\end{equation}

For realism, we consider the metric from \citet{garrido_prediction_2020} by computing the rate of structural zeros generated, but we consider the detection method of a structural zero given by \citet{darsel_robust_2025}. In this detection method, a sample is considered as a structural zero, if there is at least one combination of two of its modalities that does not exist in the data. For example, a 6-year-old individual marked as married would be considered an undesired structural zero.
This metric is named the Share of Samples with a Couple of Instances Out of Data (SSCIOD).
A more detailed derivation of this metric is given by \citet{darsel_robust_2025}.

\subsection{Comparing the initial solution over years}
\label{part:exp_static}
In our first experiments, we would like to evaluate the error due to omitting the evolving population. To that extent, we consider the three methods for generating the initial solution derived in Section \ref{part:meth_initial_sol} -- direct use of the training data, data generated by a Bayesian Network (BN), and data generated by a diffusion model. On the one hand, we generate the population based on the data from 2013 with the different methods, and we evaluate the generated population on the test set for each year between 2013 and 2021.
On the other hand, we compare the scores with populations generated with the same method, but using the training data of the same year as the evaluation data.

\begin{figure}
    \centering
    \includesvg[width=0.6\linewidth]{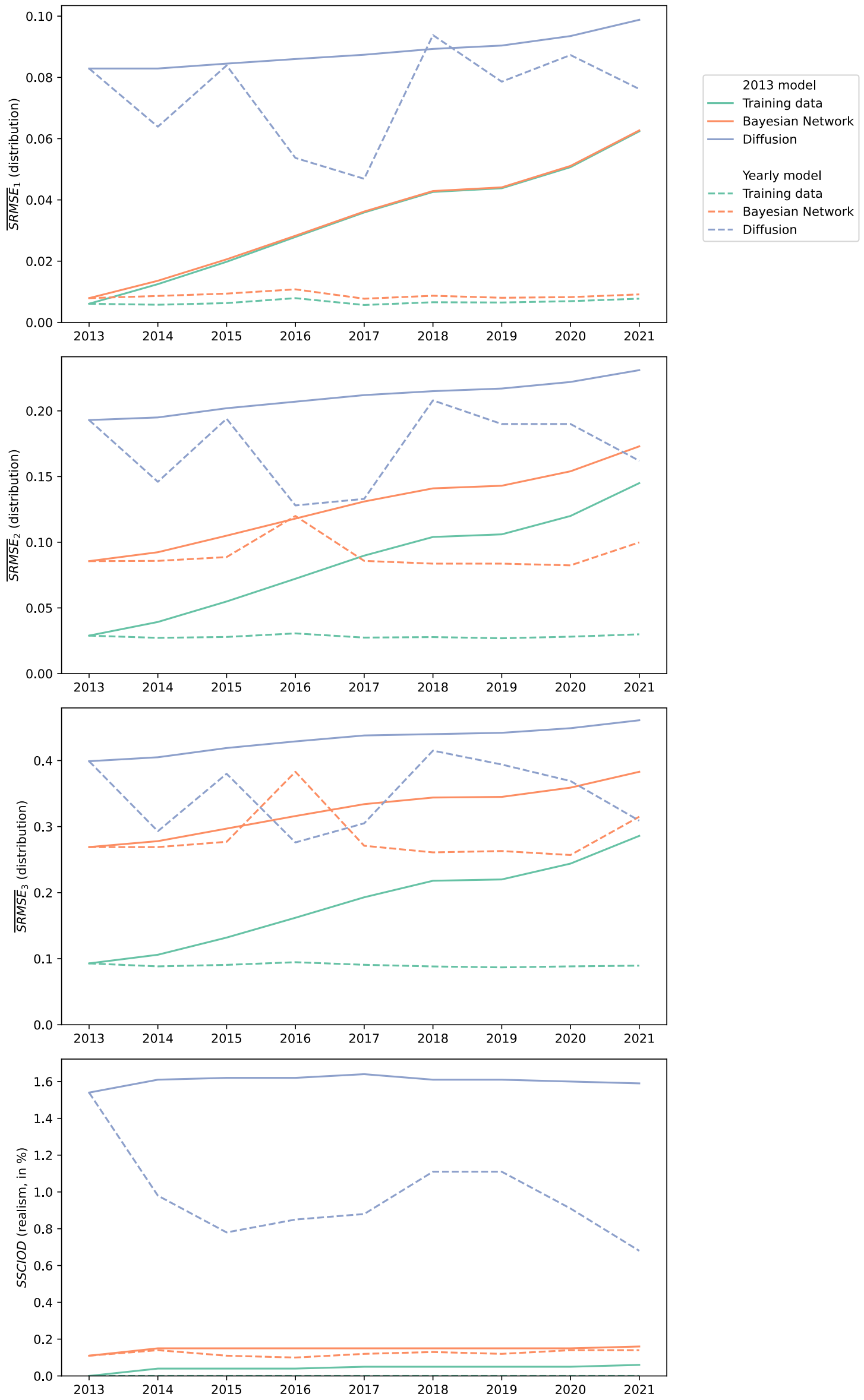}
    \caption{Comparison of the models trained with 2013 data and with yearly data over years}
    \label{fig:comparison_training}
\end{figure}
    
In Figure \ref{fig:comparison_training}, we observe the evolution of the different scores over years for the different configurations. 

The results for 2013 state that the training data performs better than the population generated with the Bayesian Network, and the diffusion model. These scores are most of the time flat, when the models are trained with yearly data, specially for training data and Bayesian Network. 

However, for the three $\overline{SRMSE}$ metrics, we observe a rise of the scores for the data generated from 2013 data over the years, whatever the generative process is. As this rise occurs only on model based on 2013 data, we deduce that it is due to a poorer representation of the population for years between 2014 and 2021, and not to the ability of the models to catch the current population. 
 For a given $\overline{SRMSE}$, the lower the initial score is, the higher the increasing is. Indeed, we observe that the scores for the training data, which has the best initial scores, rise more than the scores for diffusion models. For instance, when dealing with the marginals, the score for the training set is multiplied by $10.2$ between 2013 and 2021, but only by $1.19$ for diffusion. For the tri-variate distribution, the increase for the training set is $\times3.07$ against $\times1.15$ for diffusion.

A last observation regarding the models using 2013 data is that Bayesian Network performs very well when evaluating the distribution of the marginal distribution, comparing with the other generative model: diffusion. But, we observe that the gap with the diffusion model gets smaller over time for the tri-variate distribution. Diffusion is even better for some specific years. This makes the diffusion model interesting as it seems to be good at capturing the correlations between attributes.

The realism--measured by the $SSCIOD$--remains relatively stable for the 2013 population. This stability is expected, as a realistic individual profile should maintain its realism over time. However, we note a variation in the $SSCIOD$ for diffusion in models trained annually. These fluctuations stem from differences in the quality of the learning process and the training data. Notably, the 2013 model performs significantly worse than models from other years. 

From these first experiments, we highlight that using one year of data results in a loss of the relevance of the generated population over years. Therefore, projection methods for population synthesis are of interest, as data becomes less accurate over time. As this synthetic population is used as input for agent-based simulations, the same conclusion can be drawn about the simulations losing accuracy over time. 

\subsection{Comparing projection methods over years}

As shown in Section \ref{part:exp_static}, a synthetic population tends to be outdated after a couple of years. We aim now to evaluate the capacity of the different projection approaches--static projection (idealistic and realistic), resampling and dynamic projection--to forecast a population to overcome this issue.

\begin{figure}
    \centering
    \includesvg[width=\linewidth]{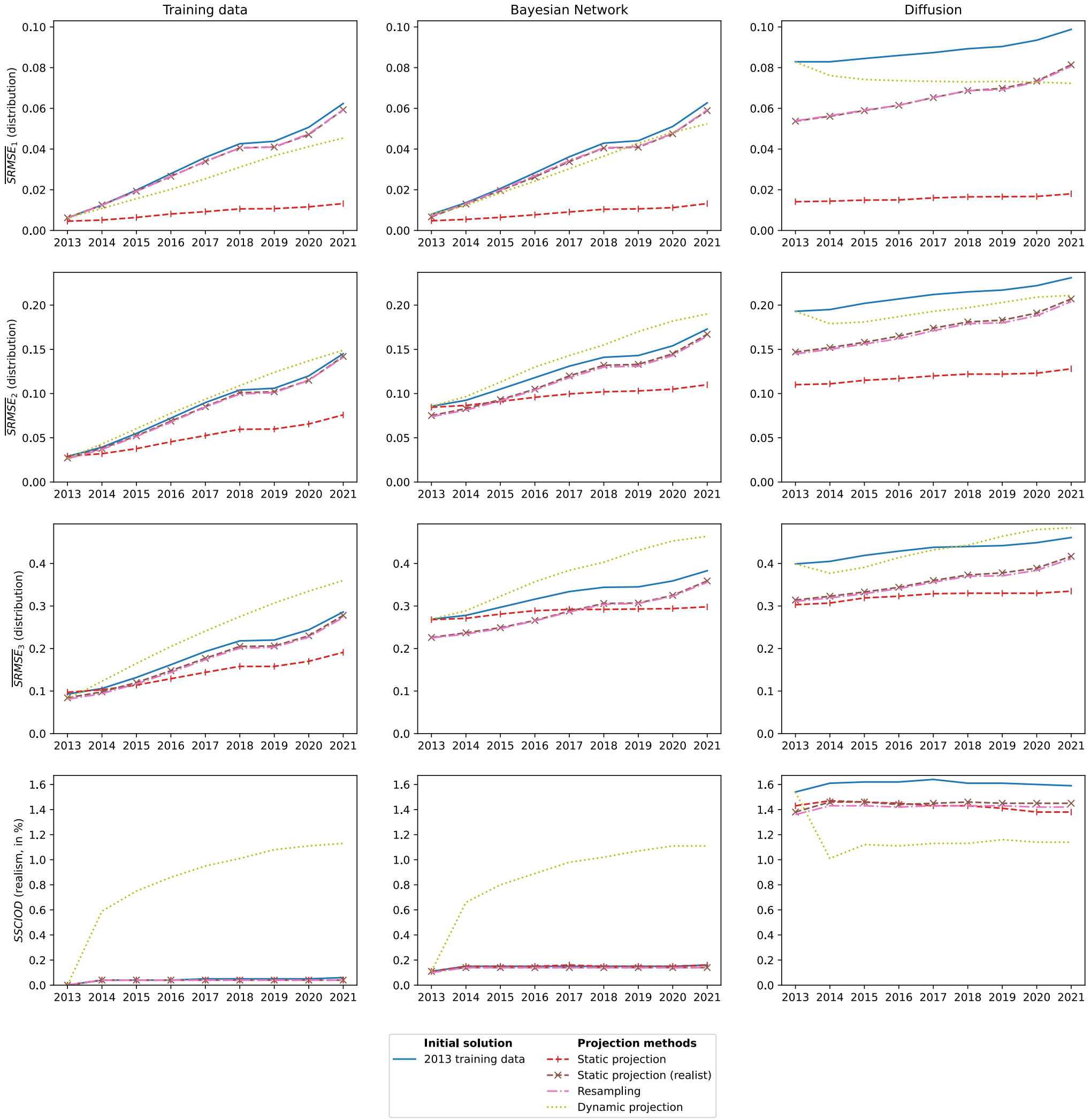}
    \caption{Comparison of the projection methods for different initial solutions}
    \label{fig:comparison_projection}
\end{figure}

Figure \ref{fig:comparison_projection} compares the different scores for the different projection methods with the different initial solutions. 

Overall, the idealistic static projection method outperforms the others, regardless of the initial population\redcomment{, as we expected}. Its performance deteriorates only slightly over time, though this decline is more pronounced when training data are used as the initial population, particularly for bi- and tri-variate statistics. The benefits of static projection are especially evident for diffusion models, which rely on the IPF steps to achieve strong scores for marginal distributions. This improvement also positively impacts bi- and tri-variate distributions. In this context, the refitting step in the static projection method significantly enhances population quality, even for the initial 2013 solution. While diffusion models initially underperform compared to other models, their results become comparable to those of the Bayesian Network across all years after refitting, and even align with training data in later experimental years.

However, these strong results are largely attributable to the richness of the projection data, as all attributes are incorporated during the fitting stage. In realistic settings, the static projection still outperforms the initial solution, but the improvement is far less substantial—particularly when using training data, where gains are minimal.

The results closely mirror those of the resampling approach, as anticipated, since both methods use the same projection data. Resampling performs slightly better due to its access to additional information through the joint distribution. The minor differences between the two methods suggest that the IPF effectively estimates the joint distribution. So in this case, the joint distribution is useless and the marginal distributions are enough. It is more valuable to have access to the marginal distributions for more attributes rather than getting the joint distribution. While the improvements from realistic static projection and resampling may seem modest, they reflect the inherent stability of the attributes considered (sex, age, and location). Since these attributes evolve slowly over time, the disparity between the initial population and updated samples remains limited.

The dynamic projection approach yields slight improvements in marginal distributions (i.e., the the $\overline{SRMSE}_1$ metric), across all three settings. However, it worsens results for other metrics compared to the baseline (the initial 2013 solution), as anticipated. Its implementation seems to disrupt the capture of more complex correlations, rendering it unsuitable in its current form. Combining the dynamic projection method with the diffusion model notably improves the realism of the generated population in the first year of the experiment.

\begin{figure}
    \centering
    \includesvg[width=0.8\linewidth]{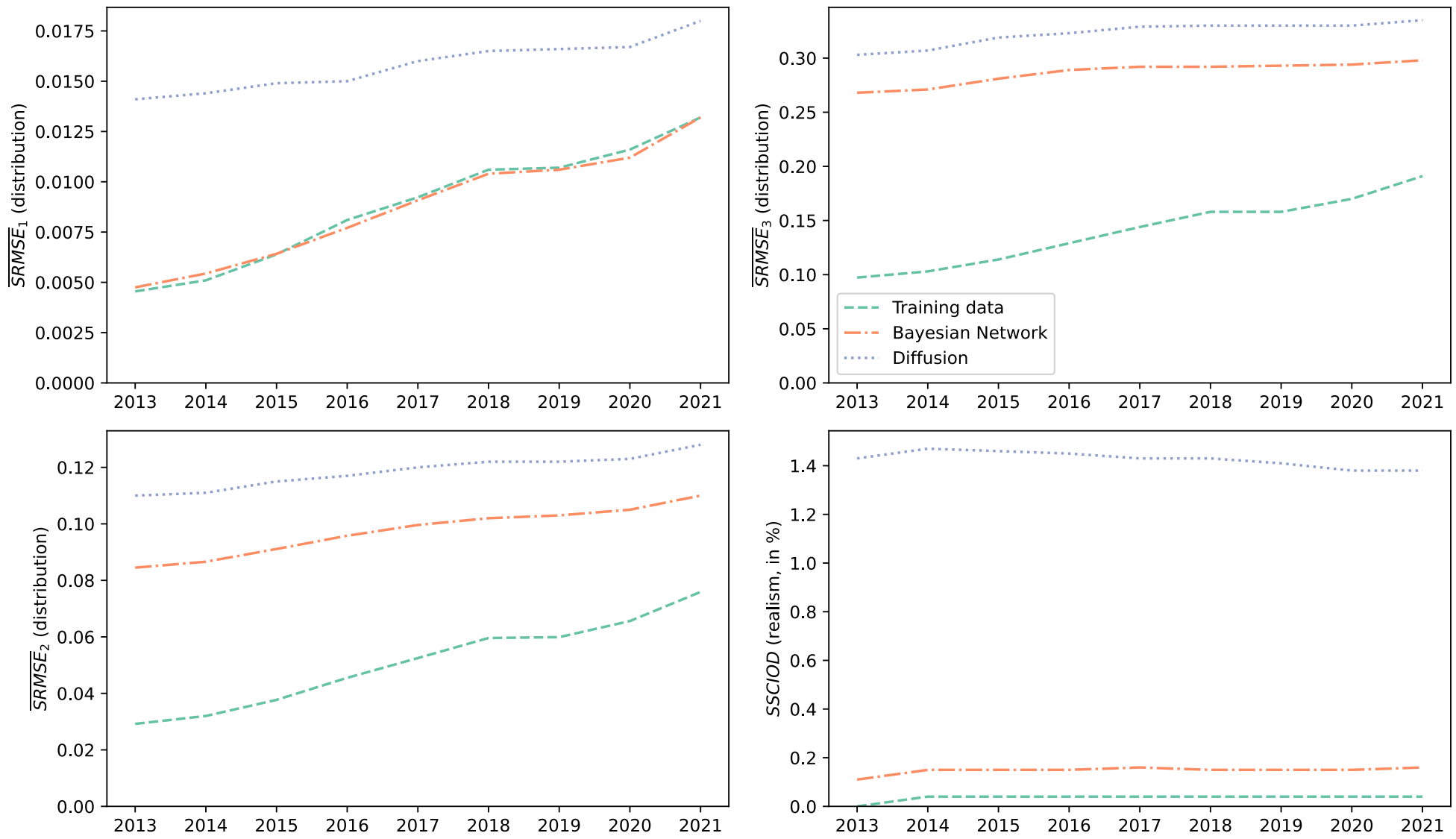}
    \caption{Comparison of the scores for static projection}
    \label{fig:comparison_static_projection}
\end{figure}

To examine the static projection method in more detail, we compare the populations fitted with this model for the various initial solutions shown in Figure \ref{fig:comparison_static_projection}. The results show that using training data consistently yields the highest scores across all metrics, with the possible exception of $\overline{SRMSE}_1$. However, the performance gap between the training data and the diffusion model gradually diminishes over time.

%To examine the idealistic static projection method in greater detail, we compare the populations fitted with this model for the different initial solutions in Figure \ref{fig:comparison_static_projection}. The results show that, with the possible exception of $\overline{SRMSE}_1$, using training data consistently yields the highest scores across all metrics. However, the performance gap between the training data and the diffusion model gradually diminishes over time.

As expected, the realism metric ($SSCIOD$) yields constant scores for methods that draw samples from a fixed pool—such as static projection and resampling—since the realism of the samples closely reflects the realism of the sampling method itself, which corresponds to the realism observed in the first year of the experiment.

% A mettre dans discussions 
% Pour Discussions ?

% Between resampling and static projection, the latter gets the best distribution scores. 
% The tri-variate $\overline{SRMSE}$ and the $SSCIOD$ for the static projection indicate that the samples are individually good, as the correlations between variables are satisfactory. Thus, the results indicates that following the refitting procedure from our static projection model is very suitable to improve the quality of a population in population synthesis, with some generative models that does not perfectly fit the marginals. 

\section{Discussions}
\label{part:discussions}
% From the experiments, it appears that fitting the population with the static projection approach is effective. In practice, static projection requires access to all marginals. This assumption can be difficult to fulfil. However, some estimations may be possible, and the algorithm could be adapted to account for the uncertainty of the values.

All our projection methods assume access to the accurate aggregated information for the target years—an assumption rarely met in practice. The dynamic projection method, for instance, requires knowledge of age-conditional margins for attributes that cannot be directly computed or derived from longitudinal data. In our experiments, the idealistic static projection method relies on 168 marginal coefficients annually, while the dynamic projection demands 3,038 coefficients. Resampling, though limited to a subset of attributes, still requires 112 coefficients yearly to capture the joint distribution.
For dynamic projection, one could consider reusing the same transition probabilities across years, as proposed by \citet{namazi-rad_generating_2014} and \citet{kukic_hybrid_2023}. However, this approach is unsuitable for static projection and resampling, as it would generate identical populations year after year.

Regarding population forecasts, \citet{wilson_methods_2022} provides an extensive review of small-area forecasting methods by age and sex. Policymakers frequently conduct such forecasts—for example, in France \href{https://www.insee.fr/fr/outil-interactif/6798992/pyramide.htm}{Age Pyramid - INSEE}) and Denmark (\href{https://www.statbank.dk/20022}{Statbank}) for their projections. Refitting with age, sex, and location aligns with our resampling model and is more feasible in practice. Thus, the realistic static projection and resampling methods are better suited for real-world applications.

For the dynamic projection method, we could have anticipated stronger performance on the  $\overline{SRMSE}_1$ metric, given its reliance on marginal frequencies and age-conditional evolution laws. However, achieving the expected margins requires an accurate age distribution fit. Our model does not account for external migrations, which can disrupt age balance over time due to age-correlated migration flows. This omission may partly explain the method’s underwhelming results.

Similarly, the idealistic static projection method does not achieve a perfect $\overline{SRMSE}_1$ score, despite optimization via the IPF algorithm. This is not due to overly lenient stopping criteria but rather to the binning process in IPF. Some attributes, such as individuals aged 110, exist in the true population but are absent from the training data, necessitating binning. This discrepancy in granularity introduces additional evaluation errors.

% Bayesian Networks do not exhibit this phenomenon because they are calibrated using observed frequencies. As a result, their marginal scores closely match those of the training data, which is expected.
% In contrast, diffusion models learn the underlying data distribution but often produce samples with imbalances. This issue stems from a balancing error rather than a structural one: the discrepancy in marginals can be mitigated, as demonstrated by static projections using 2013 data.
% One possible explanation for this imbalance is the diffusion model’s sampling process. It generates data from a Gaussian distribution, which is progressively denoised. Since Gaussian distributions are inherently non-uniform, residual imbalances may persist in the generated population.

% The weakness observed for the diffusion model with the margins in Section \ref{part:exp_static} has been solved by the static projection method. 
% Although, it does not reach the score from the training data and the Bayesian Network, the margin error is now within an acceptable range. 
With the idealistic static projection, the relative ranking of the models remains unchanged, but the gap between the diffusion model and the Bayesian Network has narrowed—particularly for multivariate errors. This highlights the power of using margins to refine the synthetic population generated from the diffusion model.
Diffusion does not generate a well balanced set, but rather a good starting solution to pick samples in, for population synthesis or agent-forecasting. 
Given the low error in the training data compared to the diffusion model’s initial error, the training data margins can serve as a reliable proxy for the true margins when the latter are unavailable. This trick could be extend to any population synthesis algorithm that struggles in respecting the margins. 
%This trick can only be performed when the generated samples are individually good. The $\overline{SRMSE}_3$ and the $SSCIOD$ give  the correlations between variables are satisfactory. Thus, the results indicates that following the refitting procedure from our static projection model is very suitable to improve the quality of a population in population synthesis, with some generative models that does not perfectly fit the marginals. 

All three forecasting methods improve the marginal distribution for the diffusion model. This improvement is intuitive for static projection and resampling, as both methods use constraints on the aggregated statistics.
For dynamic projection, however, the improvement is less immediately apparent. The enhancement emerges starting in 2014 and becomes more pronounced each subsequent year. While the marginals inform the derivation of the updating rules, they do not impose hard constraints. Nevertheless, we can mathematically justify why this method helps refitting the marginals. When marginal distributions remain relatively stable over time, i.e. when Equation \ref{eq:momentum} yields $m_k$, the error is reduced by a factor proportional to the momentum. In \ref{appendix:vanishing_error}, we provide a formal proof for cases where the momentum is uniform across all instances of a given modality.

The realism score for the diffusion model could eventually appear weak in both Figure \ref{fig:comparison_training} and \ref{fig:comparison_projection}, as it performs the worst by far. However, the share of 1\% of unrealistic samples is low compared to most generative models, what have been explored in \citet{darsel_robust_2025}. Therefore, the realism figures are acceptable in our different experiences.

Another concern regarding using the microsample as the initial solution is about privacy. As the data is directly copied for two methods, the generated population would contain directly the information about true individuals at the initial year.

% For perspectives ?

% (A moitié)
% Our resampling method is calibrated on the age, sex and department criteria. These criteria are quite general. The sex has no real impact at this scale. Departments are wide, and so the geographic criteria is not a very precise. As a result, the dynamics are quite limited at this scale. A finer geographic description could be more relevant, as aggregated statistics on less populated areas are more sensitive to dynamics.

% Comparing the initial methods, using the training data looks to be the most efficient initial solution. However, we consider a case with a microsample with 1\% of the total population. In many cases, the available data is smaller, resulting in a less representative initial solution. In our experiments, using the available data as initial solution appears to be the best solution, but it would not be necessary the case with a smaller microsample that would be less representative of the population.

\section{Conclusion \& Perspectives}
\label{part:conclusion}

This study shows that synthetic populations inevitably become outdated over time, regardless of the generative model used. This highlights the importance of using agent-forecasting techniques to keep populations up to date and ensure they are reliable for simulation purposes.

We evaluated four forecasting methods—idealistic static projection, realistic static projection, resampling, and dynamic projection—applied to a population spanning 2013 to 2021. The idealistic static projection represents an unattainable benchmark in practice but illustrates the method’s maximum potential. For dynamic projection, longitudinal data for all attributes was unavailable, as the inclusion of additional attributes (intended to better fit simulation scenarios) made such data harder to obtain. To address this, we introduced a heuristic approach based on evolving margins to generate transitions for attributes without explicit evolution laws. However, this method does not account for cross-modal interactions.

As expected, our experiments confirmed that idealistic static projection delivers the best performance, while dynamic projection fails to improve population quality when the initial solution is already robust. Resampling and realistic static projection yield similar results, suggesting that access to the joint distribution (resampling) is not essential—marginal distributions (realistic static projection) suffice to achieve comparable performance.

When comparing different models for generating the initial population—training data, Bayesian Network samples, and diffusion model samples—training data consistently produced the best results across all forecasting methods. However, the diffusion model underscored the value of a refitting step (static projection applied to the initial 2013 population), which significantly improved the quality of the generated population. Since Bayesian Networks inherently fit the training data margins, this step is unnecessary for them. We recommend performing a refitting step whenever the $\overline{SRMSE}_1$ of the generated population is substantially worse than that of the training data.

Our analysis assumes an optimistic scenario where 1\% of the total population is available—a proportion that may not reflect real-world conditions. In practice, training data are often smaller and less representative, potentially limiting the generalizability of our findings. Future research could explore the impact of smaller, less representative training datasets. Additionally, we used a broad geographical attribute (department), which lacks discriminative power. Investigating finer geographical granularity could increase forecasting complexity but may also enhance simulation accuracy.

We also operated under the idealized assumption that all aggregated statistics required for projection methods are perfectly accurate. In reality, such data are often subject to measurement errors, which could weaken or even preclude the implementation of these methods. For instance, the idealistic static projection assumes access to precise marginal distributions for all attributes at the target year—a requirement that is rarely feasible in practice.

Despite its superior performance, idealistic static projection still shows a gradual decline in multivariate distributions over time, leaving room for improvement. This method replicates samples from the initial population, effectively constraining the final distribution to remain close to the original. Consequently, attribute correlations in the final population closely mirror those in the initial population. Capturing evolving inter-attribute correlations—by modifying individuals from the initial population—represents a promising yet unexplored avenue for enhancing current methods.

Unlike \citet{kukic_adaptive_2025}, who worked at the household level, our experiments were conducted at the individual level. A household-level approach could be particularly relevant for dynamic projection, as some attributes are shared among household members or derived directly from household structure (e.g., household size). While constructing updating rules at this level requires extensive data, exploring the differences in results between individual- and household-level frameworks would be valuable.

\appendix
% \section{Derivation of transitions probabilities for not ordered attributes}
% \label{appendix:dynamic_tranistion}
% Some constraints exist in the data and should be considered when computing the transition probabilities. To avoid the generation of structural zeros, we manually create some constraints and some conditional dependencies between some attributes.

% % \begin{figure}
% %     \centering
% %     \includesvg[width=0.8\linewidth]{Illustration/Dynamic_process_diagram.svg}
% %     \caption{Diagram representing the conditionalities and constraints between attributes for the dynamic projection method}
% %     \label{fig:dynamic_projection_diag}
% % \end{figure}

% In Figure \ref{fig:dynamic_projection_diag}, the different conditional and constraint links are presented. To give a concrete example, the household type of an individual is conditionally drawn from its value the previous year, individual's age and household size. Moreover, there is a constraint from the accommodation type, in case when the individual lives in an unordinary accommodation (prison, hospital, boarding houses...). For these accommodations, no data about the household size, family position and household type is available, so a default value is set for individuals in an unordinary accommodation.

% The choice of the conditionality was done heuristically by considering only the major links, and the age to catch the evolution while ageing.

\section{Population synthesis models implementation}
\label{appendix:pop_synth}
In this appendix, we present the parameters of the generative models used for the generation of the initial solution.

\subsection{Bayesian Network implementation}
A Probabilistic Graphical Model needs to be learned first. To that extent, we use a hill-climbing approach with the BIC criterion \citep{koller_probabilistic_2009}. Then, the probabilities are computed based on the observed frequencies. In our experiments, we use the Python package called pgmpy \citep{ankan_pgmpy_2024}.

\subsection{Diffusion implementation}
\begin{figure}
    \centering
    \includesvg[width=0.9\linewidth]{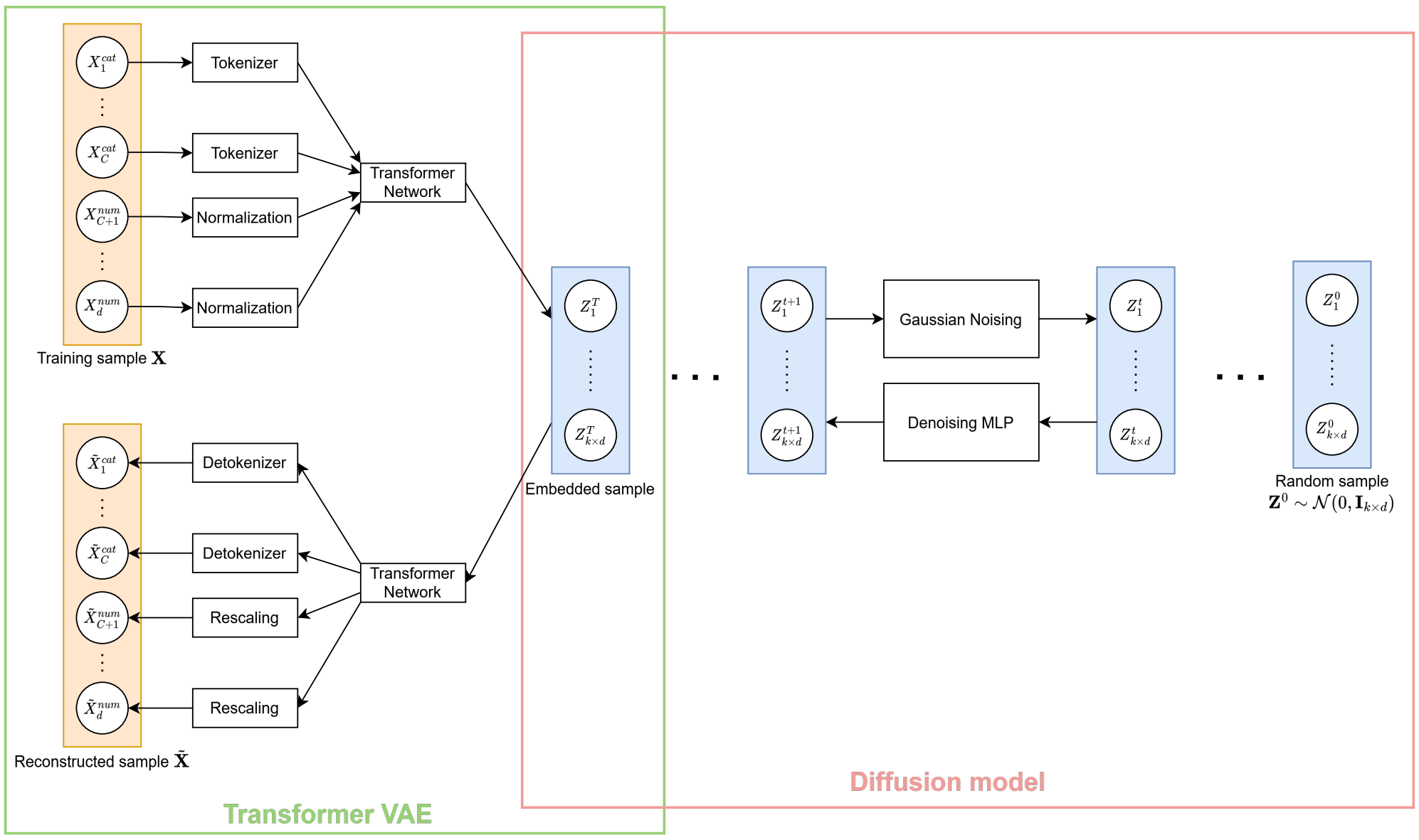}
    \caption{Architecture for the Transformer VAE network used for data embedding. $d$ is the dimension of the data. The process is slightly different for numerical and categorical variables.}
    \label{fig:tabsyn}
\end{figure}

Our implementation of diffusion is the same as \citet{darsel_robust_2025}, and is an adaptation from TabSyn model \citep{zhang_mixed-type_2023}. This model is depicted by Figure \ref{fig:tabsyn}. First, a network embeds the mixed-type data into continuous data. The embedding network is performed by a Transformer trained with a VAE structure.
Then, a diffusion model is trained on the continuous embedded data. For more detailed information on the model, we recommand to refer to the previously mentioned contributions.

\begin{figure}
    \centering
    \includesvg[width=0.4\linewidth]{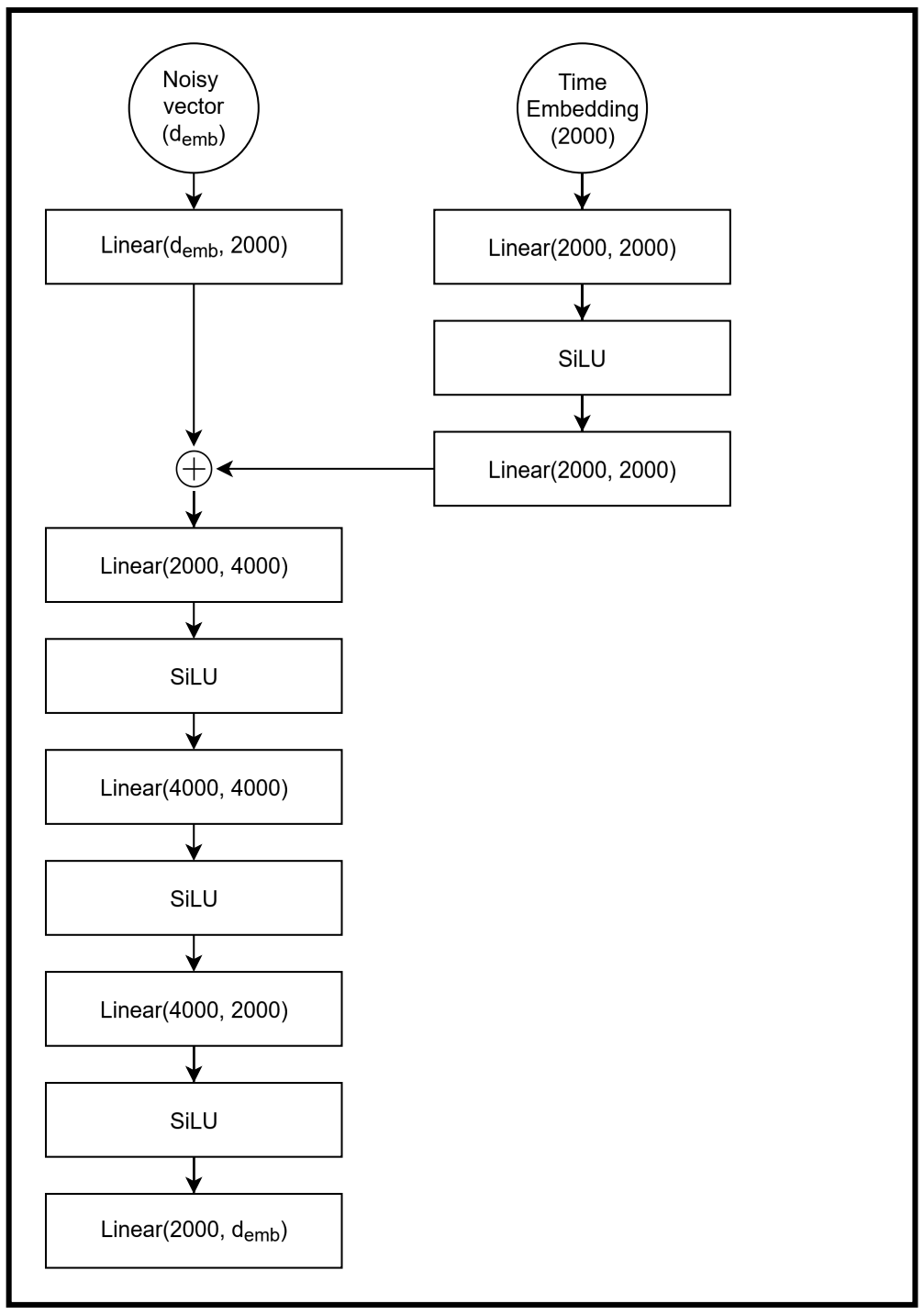}
    \caption{Architecture for the diffusion network used for data embedding. $d_{emb}$ is the dimension of the embedded data.}
    \label{fig:diffusion}
\end{figure}

The architectures for our models are given by Figure \ref{fig:transformer_VAE} and \ref{fig:diffusion}. The hyperparameters are the same used in \cite{darsel_robust_2025}.

\begin{figure}
    \centering
    \includesvg[width=0.65\linewidth]{Illustration/Transformer_VAE.svg}
    \caption{Architecture for the Transformer VAE network used for data embedding. $d$ is the dimension of the data. The process is slightly different for numerical and categorical variables.}
    \label{fig:transformer_VAE}
\end{figure}

\section{Error vanishing with momentum in dynamic projection}
\label{appendix:vanishing_error}
Here, we would like to demonstrate that an error on the margins can be reduced thanks to the momentum term. Using the notations from Section \ref{part:methodology}, we consider a modality $\omega_k$ with an error $\epsilon$ (i.e. ${}^{y}f_k(\omega_k) = \mathbbm{P}({}^{y}X_k = \omega_k) + \epsilon$). We compute the expectation for the frequency for the next year. Between two years, we assume that the changes in the margins are small enough for all modalities, so $\forall \omega\in\Omega_k, M_{\omega}=m_k\in ]0,1[$. When deriving $\mathbbm{E}[{}^{y+1}f_k(\omega_k)]$, we get:

\begin{equation}
    \begin{split}
        \mathbbm{E}[{}^{y+1}f_k(\omega_k)]=&{}^yf_k(\omega_k) \times M_{\omega_k} + (1-M_{\omega_k})\times \frac{\mathbbm{P}({}^{y+1}X_k = \omega_k)-M_{\omega_k}\mathbbm{P}({}^{y}X_k = \omega_k)}{\sum_{\omega\in\Omega_k}\mathbbm{P}({}^{y+1}X_k = \omega) - M_{\omega} \mathbbm{P}({}^{y+1}X_k = \omega)}  \\
        =&{}^yf_k(\omega_k) \times m_k + (1-m_k)\times \frac{\mathbbm{P}({}^{y+1}X_k = \omega_k)-m_k\mathbbm{P}({}^{y}X_k = \omega_k)}{\sum_{\omega\in\Omega_k}\mathbbm{P}({}^{y+1}X_k = \omega) - m_k \mathbbm{P}({}^{y+1}X_k = \omega)}  \\
        =&(\mathbbm{P}({}^{y}X_k = \omega_k) + \epsilon) \times m_k + (1-m_k)\times \frac{\mathbbm{P}({}^{y+1}X_k = \omega_k)-m_k\mathbbm{P}({}^{y}X_k = \omega_k)}{1 - m_k}  \\
        = & \mathbbm{P}({}^{y+1}X_k = \omega_k) + \epsilon \times m_k
    \end{split}
\end{equation}

The expected error is now $\epsilon \times m_k$.
%% References

%The citation must be used in following style: \cite{article-minimal} \cite{article-full} \cite{article-crossref} \cite{whole-journal}.

\bibliography{references}
\bibliographystyle{elsarticle-harv}


\end{document}